\documentclass[10.5pt,superscriptaddress,longbibliography,twocolumn]{revtex4-2}
\usepackage{textcomp}
\usepackage{times}
\usepackage{graphicx}
\usepackage{float}
\usepackage{latexsym,amsmath,amssymb,bm,euscript}
\usepackage{color}
\usepackage{subfigure}
\usepackage{epstopdf}
\usepackage[colorlinks=true,linkcolor=blue,citecolor=blue,urlcolor=blue]{hyperref}
\usepackage{hyperref}
\usepackage{soul}
\usepackage[normalem]{ulem}
\usepackage{mathrsfs}
\usepackage{amsmath,amssymb}
\usepackage{lettrine}
\usepackage{xfrac}
\usepackage{xspace}
\usepackage{textcomp}
\usepackage{wasysym} 
\usepackage{xr}

\newcommand{\Eq}[1]{Eq.~\eqref{#1}}

\newcommand{\Fig}[1]{Fig.~\ref{#1}}

\newcommand{\Tab}[1]{Tab.~\ref{#1}}

\graphicspath{{./Figs/}}
\begin{document}

\title{Realization of Topological Mott Insulator in a Twisted Bilayer Graphene Lattice Model}

\author{Bin-Bin Chen}
\affiliation{School of Physics, Beihang University, Beijing 100191, China}
\affiliation{Department of Physics and HKU-UCAS Joint Institute of Theoretical and Computational Physics, 
	The University of Hong Kong, Pokfulam Road, Hong Kong, China}

\author{Yuan Da Liao}
\affiliation{Beijing National Laboratory for Condensed Matter Physics, and Institute of Physics, 
Chinese Academy of Sciences, Beijing 100190, China}
\affiliation{School of Physical Sciences, University of Chinese Academy of Sciences, Beijing 100190, China}

\author{Ziyu Chen}
\affiliation{School of Physics, Beihang University, Beijing 100191, China}

\author{Oskar Vafek}
\affiliation{Department of Physics, Florida State University, Tallahassee, FL 32306, USA}
\affiliation{National High Magnetic Field Laboratory, Tallahassee, Florida, 32310, USA}

\author{Jian Kang}
\email{jkang@suda.edu.cn}
\affiliation{School of Physical Science and Technology \& Institute for Advanced Study, Soochow University, Suzhou, 215006, China}

\author{Wei Li}
\email{w.li@buaa.edu.cn}
\affiliation{School of Physics, Beihang University, Beijing 100191, China}
\affiliation{CAS Key Laboratory of Theoretical Physics, Institute of Theoretical Physics, Chinese Academy of Sciences, Beijing, 100190, China}

\author{Zi Yang Meng}
\email{zymeng@hku.hk}
\affiliation{Department of Physics and HKU-UCAS Joint Institute of Theoretical and Computational Physics, 
The University of Hong Kong, Pokfulam Road, Hong Kong, China}

\begin{abstract}
\end{abstract}

\date{\today}
\maketitle


\noindent\textbf{Magic-angle twisted bilayer graphene has recently become a {thriving} 
material platform realizing correlated electron phenomena taking place within 
its topological flat bands. Several numerical and analytical methods have been 
applied to understand the correlated phases {therein}, revealing some similarity 
with the quantum Hall physics. In this work, we provide a Mott-Hubbard perspective 
for the TBG system. Employing {the} large-scale density matrix renormalization group 
on the lattice model containing the projected Coulomb interactions only, we identify 
a first-order quantum phase transition between the {insulating} stripe phase and the 
quantum anomalous Hall state with the Chern number of $\pm 1$. Our results not 
only shed light on the mechanism of {the quantum anomalous Hall state} 
discovered at three-quarters filling, but {also} provide an example of the topological 
Mott insulator, i.e., the {quantum anomalous Hall} state in the strong coupling limit.
}\\


\bigskip\noindent\textbf{Introduction}

\noindent Twisted bilayer graphene (TBG) burst on the scene as a tunable two carbon-atom layers thick system realizing a remarkable multitude of interaction-driven macroscopic quantum phenomena~\cite{cao2018correlated, cao2018unconventional, Yankowitzeaav2019,lu2019superconductors,xie2019spectroscopic,sharpe2019emergent,serlin2020intrinsic,stepanov2019interplay,kerelsky2019maximized,
jiang2019charge,choi2019,Nuckolls_2020,Uri_2020,choi2020tracing,wu2020chern,Saito_2021,das2020symmetry,park2020flavour,zondiner2019cascade,wong2019cascade}. Although significant progress has been achieved in understanding the nontrivial topology of the narrow bands, 
as well as the correlated electron states 
in the magic-angle TBG~\cite{bistritzer2011moire, kang2018symmetry, Po2018, koshino2018maximally, po2018origin, kang2019strong,
liu2019correlated,kang2020nonabelian,xie2018nature,ahn2019failure,
fragile_topology,LiuValley2019,wu2020collective,zhang2020correlated,liu2018pseudo,tarnopolsky2019origin,Carr2019Exact,Ren2021,guinea2018electrostatic,Guinea2019Continuum,yuan2018model,xu2018topo},
many important questions remain open. 
One of the most fascinating question is the origin 
and the mechanism of the quantum anomalous Hall (QAH) 
state with Chern number $C= \pm1$~\cite{sharpe2019emergent,serlin2020intrinsic} 
at three-quarters filling of the system, aligned with the hexagonal boron nitride (hBN), 
and the insulating state which replaces the QAH in devices without the hBN alignment.

Currently, the prevailing opinion is that the QAH can be obtained from narrow band models with large Coulomb interactions~\cite{bultinck2019ground,kang2020nonabelian,soejima2020efficient,lian2020tbg,Kwan2021}, but that the nontrivial topology of the narrow bands prevents a faithful construction of local ``Hubbard-like'' tight binding models that locally respect all the symmetries~\cite{Po2018}. Although there exists no a priori Wannier obstruction, as the narrow bands' total Chern number vanishes, there is yet no clear understanding of how the QAH could arise within such correlated lattice model, even in principle, in the limit where the Coulomb interactions dominate the kinetic energy.

\begin{figure*}[t!]
\includegraphics[angle=0,width=0.88\linewidth]{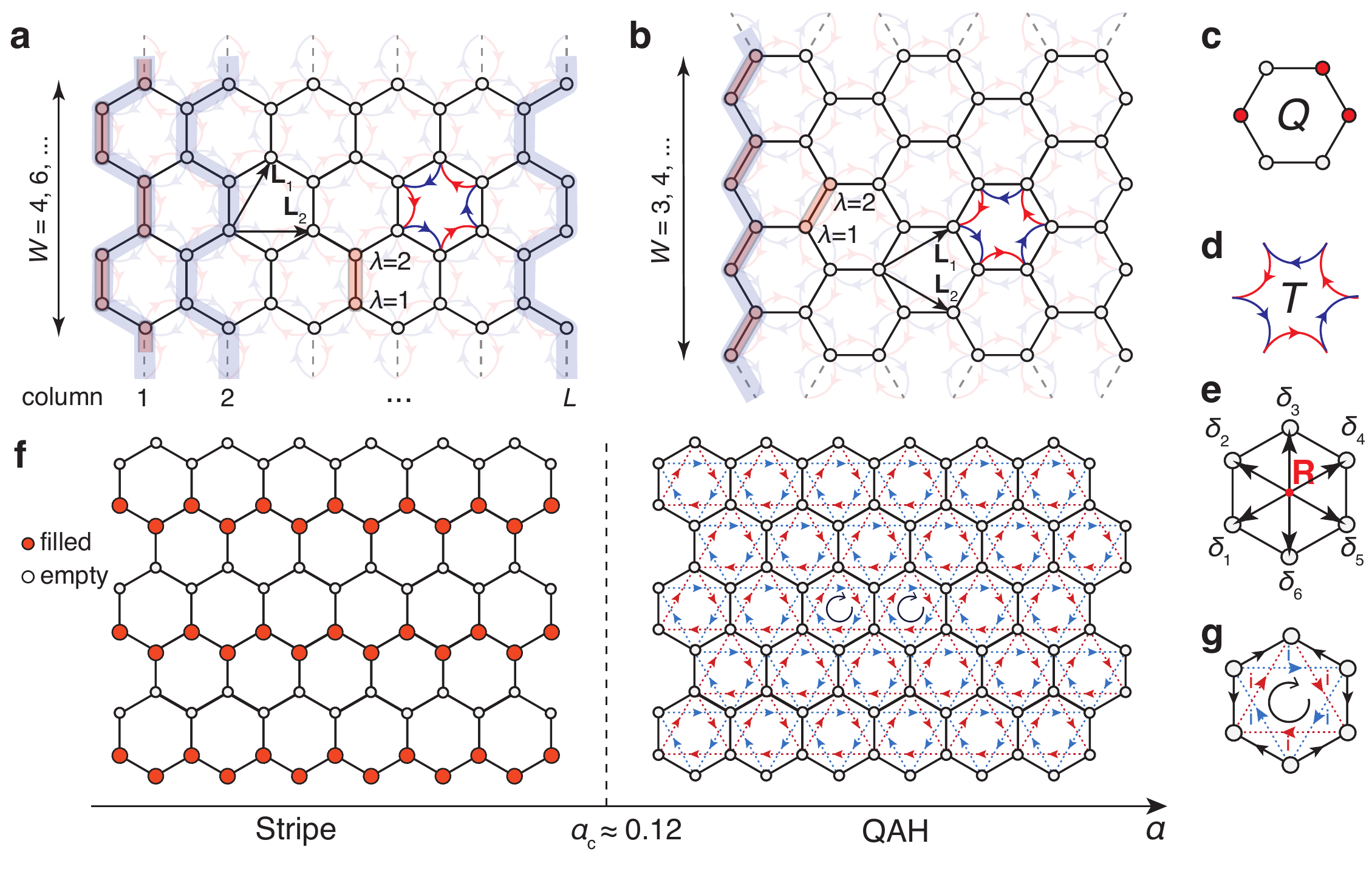}
\caption{\textbf{The Honeycomb Moir\'e Lattice Model and Phase Diagram.} 
\textbf{a} YC and \textbf{b} XC geometries with PBC along 
vertical ($\mathbf{L}_1 - \mathbf{L}_2$ for XC and $2\mathbf{L}_1 - \mathbf{L}_2$ for YC) 
and OBC along horizontal direction. 
The number of sites on the cylinders is $N =W\times L\times 2$,
with length $L$ (the number of vertical armchair/zigzag chains, c.f. the grey-shaded lines)
and $W$ is the number of 2-site unit cells (c.f. the red-shaded rectangles) along those chains.
\textbf{c} shows the cluster charge operator $Q_{\varhexagon}$, 
which counts the electron number in a hexagon 
and \textbf{d} demonstrates the assisted hopping term $T$ with alternating-sign structure. 
\textbf{e} The labeling of six sites within hexagon $\mathbf{R}$. 
\textbf{f} The phase diagram contains two distinct 
insulating phases, i.e., the stripe phase for $\alpha<\alpha_c$, 
and the QAH state for $\alpha>\alpha_c\simeq0.12$. 
\textbf{g} The schematic plot of the emergent current through
{a mean-field tight-binding analysis of the QAH state}.
}
\label{Fig:Model}
\end{figure*}

Precisely such a state was sought by Raghu, Qi, Honerkamp and  S.-C. Zhang in an entirely different context~\cite{topMott2008}, coining the term topological Mott insulator (TMI), which we define to be a QAH in a strong coupling limit of a local lattice model with a vanishing ratio of the bandwidth 
to Coulomb interaction. 
However, the original proposal~\cite{topMott2008} was subsequently shown not to host a QAH, and therefore not TMI either~\cite{Jia2013,Capponi2015}.  More recent works have found the interaction-induced QAH state in a different model, 
but it is stabilized by the kinetic energy and necessitates sizable bandwidth~\cite{Sun2009,Zhu2016,Gong2018}. 
Because it gives way to more conventional Mott insulators in the strong coupling regime~\cite{Gong2018}, these models do not host a TMI.

Here we show that the TMI is realized in a simple lattice model introduced by two of the authors
as a local description of the correlations within the TBG narrow bands~\cite{kang2019strong,YDLiao2020,YDLiao2020review}. 
The key new ingredients are the off-site terms appearing alongside the usual on-site terms in the projected density operator.
Physically, such terms originate in the extended multi-peak nature of the maximally localized Wannier states~\cite{kang2018symmetry,koshino2018maximally} arising from the nontrivial topology~\cite{po2018origin,liu2018pseudo,song2019all,po2018origin,khalaf2021charged,Zou2018,Xie2020Topology} of the narrow bands, and, importantly, remain finite even when the bandwidth vanishes. 

\bigskip\noindent\textbf{Results}

\noindent\textbf{Honeycomb Moir\'e Lattice Model.}  
In the strong coupling limit, the aforementioned model 
(as illustrated in the upper 
panels of Fig.~\ref{Fig:Model}) is
\begin{equation}\label{Eq:H}
 H =  U_0\sum_{\varhexagon}({Q_{\varhexagon}} 
 + \alpha T_{\varhexagon} - 1)^2, 
\end{equation} 
where $U_0$ constitutes the overall energy scale 
in the problem ($\approx 40$~meV in TBG and 
set to unity henceforth).
$Q_{\varhexagon} \equiv \frac{1}{3}\sum_{l=1}^6 c^\dagger_{{\bf R}+\delta_l} 
c^{\phantom{\dagger}}_{{\bf R}+\delta_l}$ represents 
the cluster charge term~\cite{po2018origin,xu2018kekule,kang2018symmetry,YDLiao2019,YDLiao2020,YDLiao2020review}  
[c.f. \Fig{Fig:Model}(c)], and 
$T_{\varhexagon} \equiv \sum_{l=1}^{6}
[(-1)^l c^\dagger_{{\bf R}+\delta_{l+1}}c^{\phantom{\dagger}}_{{\bf R}+\delta_l} + h.c.]$ 
represents the Coulomb induced hopping with alternating sign 
[c.f. \Fig{Fig:Model}(d)]. Fermion annihilation and creation operators
$c_{{\bf R}+\delta_l}^{\,}$ and $c_{{\bf R}+\delta_l}^{\dagger}$ are defined 
at the sites of the honeycomb lattice ${\bf R}+\delta_l$, 
where ${\bf R}=m_1{\bf L}_1+m_2{\bf L}_2$ with integer $m_{1,2}$ 
spans the triangular Bravais lattice. The hexagon centers, over which we sum 
in Eq.(\ref{Eq:H}), are connected to the six nearest honeycomb lattice
sites $l = 1, 2, \cdots 6$ through $\delta_{l}$ [c.f. Fig.~\ref{Fig:Model}(e)]. 
As we focus on the three-quarters filling of the TBG, 
where the spin and orbital degrees of freedom are 
assumed to be polarized, Eq.~(\ref{Eq:H}) thus constitutes
a simplification to the full Hamiltonian of Ref.~\cite{kang2019strong}. 
The parameter $\alpha$ controls the relative strength of 
	charging and assisted-hopping of the projected Coulomb interaction. 
	It originates from the overlap of two neighboring Wannier states in the continuum 
	model and thus depends on the lattice relaxation. Due to the background charge from the remote bands, which is approximated to be uniform in Eq.~(\ref{Eq:H}), the projected Coulomb interaction is in the form of density-density repulsion~\cite{bultinck2019ground,VafekKang2020,bernevig2020tbg}, 
	instead of being normal ordered.
	{Although the projected interaction contains other terms such as next-nearest neighbor interaction, the more detailed calculations at the chiral limit have shown that the interaction induced dispersion of the charged excitation at the charge neutrality point is dominated by $\alpha$, the nearest neighbor assisted hopping~\cite{VafekKang2021Prep}.  } 

\begin{figure}[htb]
\includegraphics[angle=0,width=1\linewidth]{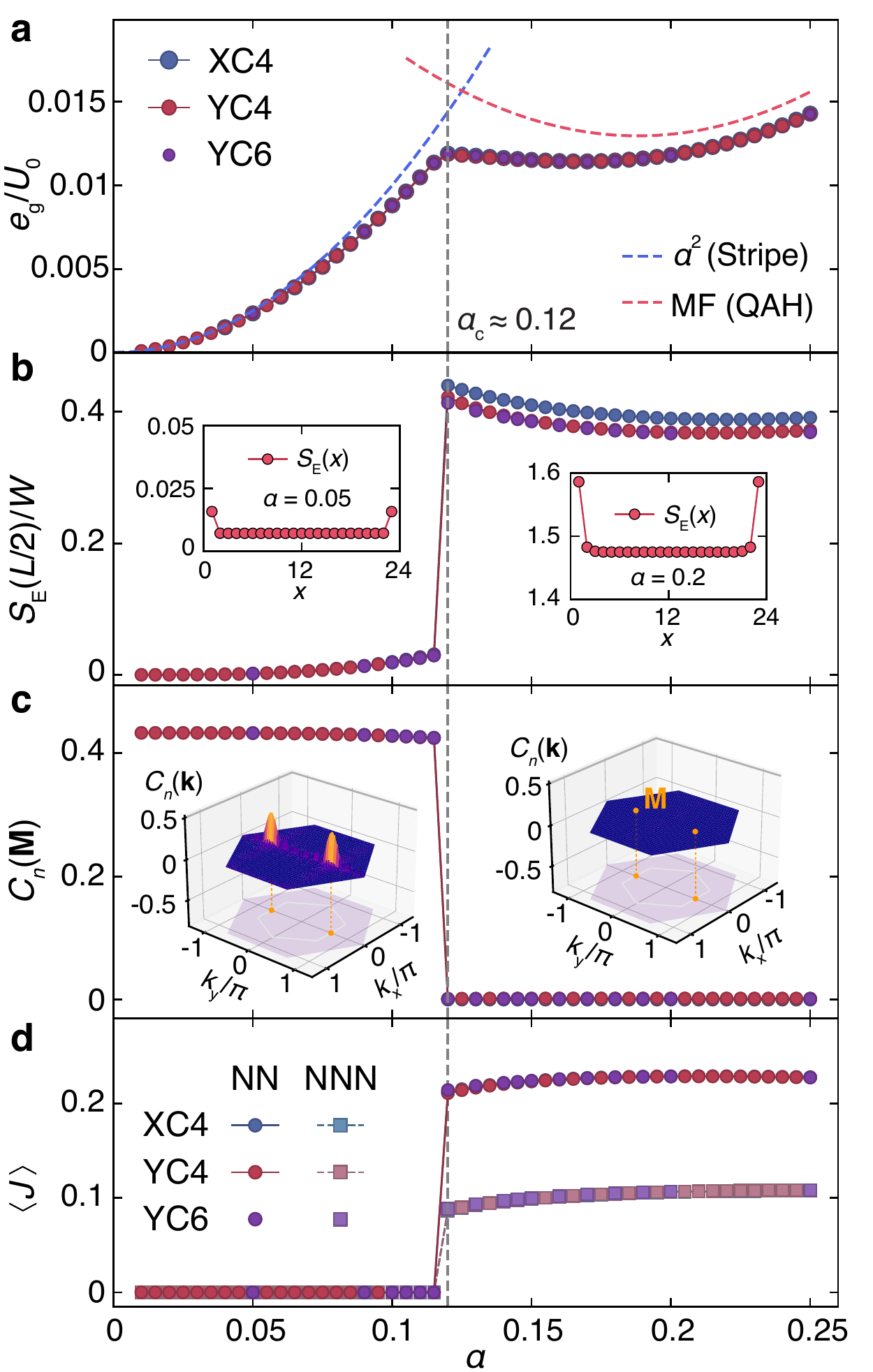}
\caption{\textbf{Identification of two insulating phases.}
\textbf{a} The ground-state energy per site 
$e_g \equiv \frac{1}{N}\langle \psi_g | \hat H | \psi_g\rangle$, 
shown as a function of $\alpha$, with total number of sites 
$N=2WL$ and $|\psi_g\rangle$ the DMRG ground state. 
\textbf{b} Entanglement entropy $S_E$,  
\textbf{c} stripe order parameter $C_n({\bf M})$, 
\textbf{d} both {correlations} $\langle J\rangle_\mathrm{NN}$ 
and $\langle J\rangle_\mathrm{NNN}$, 
are shown versus $\alpha$, all showing abrupt 
changes of behavior at $\alpha_c\simeq0.12$. 
The mean-field energies for both phases are as well shown in 
panel \textbf{a}. The detailed entanglement profile $S_E$ vs. 
subsystem $x$ is shown in the inset of (b),
and $C_n({\bf k})$ vs. ${\bf k}$ in the first Brillouin zone (BZ) 
shown in the inset of (c).
}
\label{Fig:Phases}
\end{figure}

The original bandwidth $W \sim 8$meV~\cite{koshino2018maximally} is much smaller than $U_0$, suggesting the system is in the strong coupling regime. Furthermore, after the states on the remote bands are integrated out, the superexchange interaction ($\lesssim 5\times 10^{-3} e^2/(\epsilon L_m)$) is found to be negligible compared with the projected Coulomb interaction~\cite{VafekKang2020}; this justifies neglecting additional fermion bilinear (kinetic) terms in Eq.~(\ref{Eq:H}). {The kinetic term, as well as the further-range assisted hopping terms, may shift the critical value $\alpha_c$ of the phase transition but do not qualitatively change the phase diagram in Fig.~\ref{Fig:Model}(f). }
In addition, we do not include the additional symmetry breaking term produced by the possible hBN alignment that favors the QAH phase~\cite{ZaletelQAH}, {but focus on the topological phase transitions purely driven by interactions}.

It is worth emphasizing that Eq.~\eqref{Eq:H} corresponds to the leading order terms when the distance 
to the gates $l_g$ is about the same as the moir\'e lattice constant $|\bf L_1|$, 
and thus the electron-electron repulsion decays exponentially when the inter-electron separation is larger than $|\bf L_1|$~\cite{kang2019strong}. 
With larger $l_g$, the longer range aspect of the Coulomb repulsion will have to be included, 
but because currently there is no experimental indication that there are significant changes in the nature the insulating 
states for different $l_g$ ~\cite{cao2018correlated,Yankowitzeaav2019,Liu2021tuning}, 
it is reasonable to neglect the longer range terms in Eq.~\eqref{Eq:H}. 
{We should note that terms in Eq.~\eqref{Eq:H} are purely real, and because  the two QAH states with opposite 
	Chern numbers transform into each other under complex conjugation, the QAH state is not \textit{a priori} favored by this model.}
In what follows, we will demonstrate that, for a range of $\alpha$, Eq.~\eqref{Eq:H} naturally leads to the topological Mott insulator ground state via spontaneous symmetry breaking without including any other interactions or kinetic terms.


\bigskip\noindent\textbf{Phase Diagram.} 
We solve the TBG lattice model in Eq.~(\ref{Eq:H}) 
using DMRG on long cylinders of XC [zigzag, Fig.~\ref{Fig:Model}(a)] 
and YC [armchair, Fig.~\ref{Fig:Model}(b)] 
geometries, with widths $W$ up to 6 and lengths $L$ up to 24. 
The details of DMRG implementation 
and finite-size analysis are given in the Methods and 
Supplementary Note 1. 
The obtained ground state phase diagram,
as a function of $\alpha$, 
is shown in \Fig{Fig:Model}(f). We identify two gapped
insulating phases: a stripe phase with charge density 
wave (CDW) for small $\alpha$, and a TMI 
phase for $\alpha > \alpha_c\approx 0.12$. 
These two ground states are separated by 
a first-order quantum phase transition (QPT).
In \Fig{Fig:Phases}, we show results for various quantities,
including the ground state energy $e_g$, entanglement 
entropy $S_E$, charge structure factor $C_n$, and 
{the imaginary part of the equal time 
correlation $\langle J\rangle \equiv \frac{\textit{i}}{2} \langle
(c_{l}^\dagger c^{\,}_{l'} - c_{l'}^\dagger c^{\,}_{l})\rangle$.} 
As shown in \Fig{Fig:Phases}(a), 
the $e_g$ curve exhibits a discontinuity 
in the slope (a kink) at $\alpha_c$, 
indicating the first-order QPT.
In \Fig{Fig:Phases}(b), we calculate the entanglement entropy 
$S_E(x) \equiv -\mathrm{Tr}[\rho_\mathcal{A}(x)
\ln(\rho_\mathcal{A}(x))]$, with $\rho_\mathcal{A}(x)$ 
the reduced density matrix of the subsystem 
$\mathcal{A}$ consisting of the first $x$ columns
[c.f. \Fig{Fig:Model}(a,b)]. 
By setting $x=L/2$ (for even $L$), i.e., 
cutting at the very center of the system, 
we compute $S_E(L/2)$ and show it vs. $\alpha$ 
in \Fig{Fig:Phases}(b), where an evident ``jump" 
takes place right at the QPT. 
In addition, for $\alpha<\alpha_c$, the negligibly small 
$S_E(L/2)$ indicates the existence of a nearly direct product 
state with virtually no charge fluctuations in the CDW pattern.
On the other hand, the sizable $S_E(L/2)$ 
for $\alpha>\alpha_c$ indicates
a finite amount of quantum entanglement in the ground state.
In the insets of \Fig{Fig:Phases}(b),
$S_E(x)$ vs.~subsystem length $x$ shows a flat plateau 
in the bulk of the system, indicating that both phases in 
Fig.~\ref{Fig:Model}(f) are gapped, consistent with the
exponentially decaying single-particle Green's functions
also obtained by our DMRG (see the Supplementary Note 1).

\bigskip\noindent\textbf{Stripe and QAH Insulators.} 
The emergence of the stripe phase at small $\alpha$ can be 
understood from a perturbative analysis~\cite{kang2019strong}. 
Up to second-order corrections (c.f. Supplementary Note 2), 
we find the ground-state energy $e_g/U_0 \simeq \alpha^2$, 
and plot it together with the DMRG results in 
Fig.~\ref{Fig:Phases}(a), where the high 
accuracy of this analytical calculation 
can be clearly seen. The CDW order can be characterized by 
the structure factor, $C_n({\bf k}) \equiv  \frac{1}{N} \sum_{\lambda=1}^2 \sum_{\bf R} 
e^{-i{\bf k}\cdot (\bf R + \delta_{\lambda})} \tilde{n}_{\bf R,\lambda}$, 
where the quantity
$\tilde{n}_{\bf R, \lambda} = \langle c_{\mathbf{R} + \delta_\lambda}^\dagger 
c_{\bf{R} + \delta_\lambda}^{\phantom{\dagger}} \rangle -1/2$ counts
the number of electrons (with respect to the half filling) 
on the honeycomb site $\bf{R} + \delta_\lambda$. 
In \Fig{Fig:Phases}(c), we find that $C_n({\bf k})$ peaks at 
${\bf M}=(0,\frac{2\pi}{\sqrt{3}|\mathbf{L}_1|})$ for $\alpha < \alpha_c$, 
and drops abruptly to 0 for $\alpha > \alpha_c$, confirming that the small-$\alpha$ regime has a CDW order, while for $\alpha > \alpha_c$ the insulating phase has no charge order. Remarkably, this $\alpha > \alpha_c$ regime turns out to be a topological phase with spontaneous time-reversal symmetry (TRS) breaking and a quantized Hall conductance, i.e., a QAH phase.

\begin{figure}[htb]
\includegraphics[angle=0,width=\linewidth]{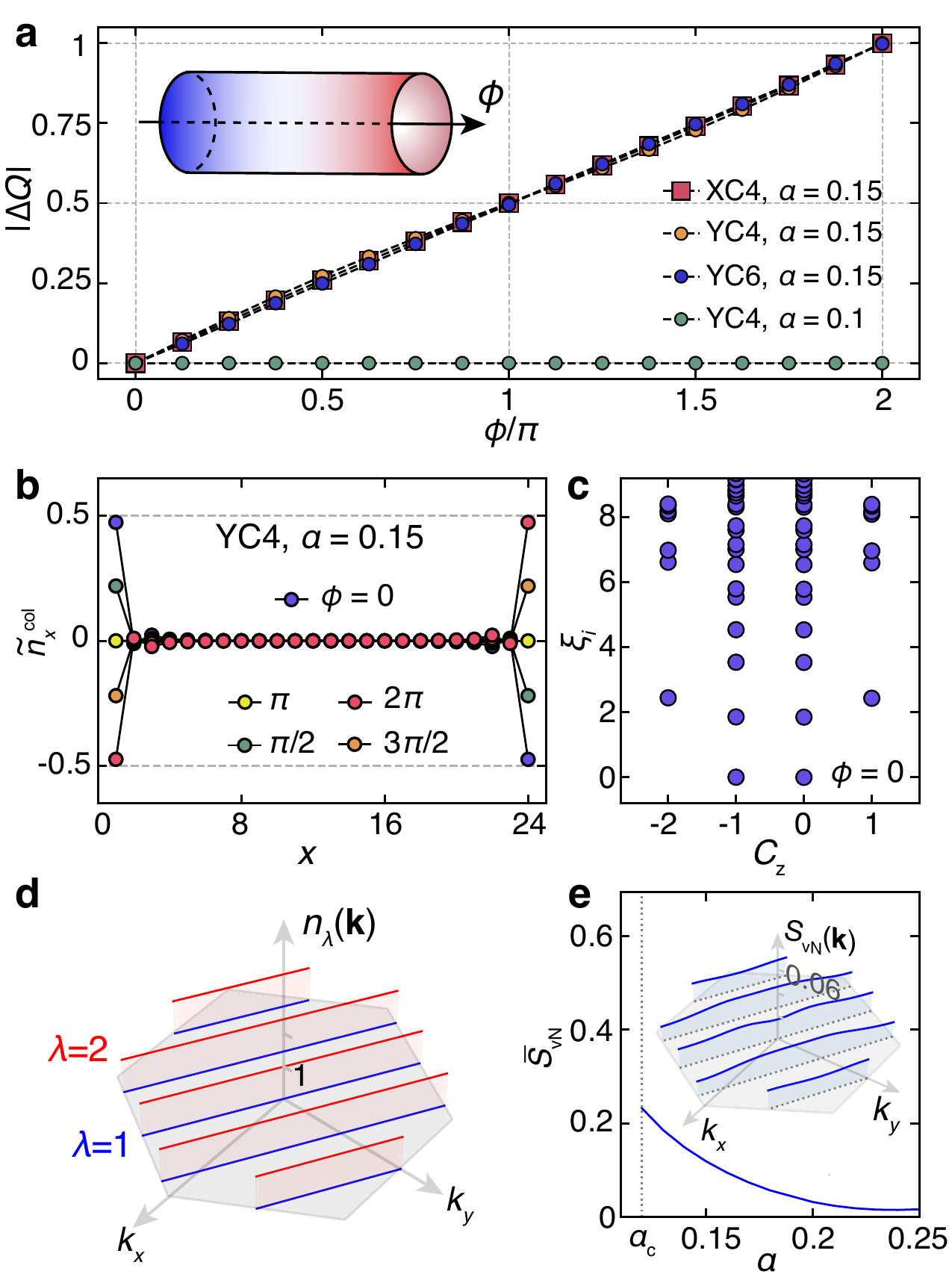}
\caption{\textbf{Quantized Hall Conductance and QAH State.}
In the systems of both width $W=4,6$, 
a flux $\phi\in[0,2\pi]$ is threading through the cylinder, 
\textbf{a} One electron is pumped from one edge 
to the other for $\alpha=0.15$ (QAH phase), 
while no charge response is observed for $\alpha=0.1$ (stripe phase).
\textbf{b} In the real-space charge distribution of YC4 cylinder, 
no accumulations are observed in the bulk, i.e., 
only the charge near the left edge is pumped. 
\textbf{c} Entanglement spectrum computed at the central bond 
(between two columns) shows a two-fold degeneracy. 
For a typical QAH state with $\alpha=0.25$, we show in 
\textbf{d} the charge density $n_\lambda(\bf k)$, with $\lambda$ labeling 
the two eigenvalues of the $2\times2$ $\tilde{G}(\bf k)$ 
matrix associated with two sublattices. 
In \textbf{e} the von Neumann Entropy $\bar{S}_\mathrm{vN}$ 
averaged over all the $\bf k$ points, is shown versus $\alpha$,
where the $S_\mathrm{vN}$ distribution in BZ 
is shown in the inset (with also $\alpha=0.25$).}
\label{Fig:Flux}
\end{figure}

To reveal the TRS breaking in the large-$\alpha$ QAH phase, 
{in Fig.~\ref{Fig:Phases}(d) we show the correlation $\langle J\rangle$} 
on both the nearest-neighbor (NN) and next-nearest-neighbor (NNN) $(l,l')$ pairs. 
We find a finite value of $\langle J\rangle_\mathrm{NN}\sim0.22$ 
and $\langle J\rangle_\mathrm{NNN}\sim0.1$
in the bulk of the cylinder for large-$\alpha$ phase, 
while they vanish in the stripe phase. 
{In the QAH phase, the real part of 
$\langle c_l^\dag c_{l'}^{\phantom{\dag}}\rangle$ is 
negligibly [$O(10^{-7 \sim -8})$] smaller compared to its imaginary part, 
and thus $\langle c_l^\dag c_{l'}^{\phantom{\dag}}\rangle$ 
emerging from interactions is virtually purely imaginary.} 
The corresponding hopping process thus acquires a $\pi/2$ phase 
[labeled as $\textit{i}$ in Fig.~\ref{Fig:Model}(g)], 
rendering a $3\pi/2$ flux for a circulating triangular loop current, 
which resembles the Haldane model~\cite{Haldane1988}.  
The difference is that the TRS breaking NNN hopping term 
is introduced explicitly in the Haldane model, while here it
emerges spontaneously due to electron interactions, a typical 
feature of topological Mott insulators. We also note that in 
a recent quantum Monte Carlo simulation applied at charge 
neutrality~\cite{YDLiao2020} (i.e. even integer filling), a quantum valley Hall state is found at intermediate coupling for a specific choice of kinetic energy terms. Such a state is different from the QAH found at odd integer filling here as it preserves the TRS with helical valley edge modes and undergoes a first order phase transition into intervalley coherent insulator at strong coupling, consistent with the exact results obtained in Ref.~\cite{kang2019strong}.

\bigskip\noindent\textbf{Quantized Hall Conductance.}
To reveal the topological properties in the large-$\alpha$ phase, 
we perform a flux insertion experiment on the cylindrical
geometry [c.f.~the inset of \Fig{Fig:Flux}(a)] and 
compute the Hall conductance. We thread a $\phi$-flux 
along the cylinder by modifying the boundary condition 
$c_{\mathbf{R} + W(\mathbf{L}_1 - \mathbf{L}_2) + \delta_{\lambda}} 
\equiv c_{\mathbf{R} + \delta_{\lambda}}$ to $c_{\mathbf{R} + W(\mathbf{L}_1 - \mathbf{L}_2) + \delta_{\lambda}} \equiv e^{-i \phi}  c_{\mathbf{R} + \delta_{\lambda}}$ for XC geometry and $c_{\mathbf{R} + W(\mathbf{L}_1 - \mathbf{L}_2/2) + \delta_{\lambda}} \equiv c_{\mathbf{R} + \delta_{\lambda}}$ to $c_{\mathbf{R} + W(\mathbf{L}_1 - \mathbf{L}_2/2) + \delta_{\lambda}} \equiv e^{-i \phi}  c_{\mathbf{R} + \delta_{\lambda}}$ for YC geometry. During the process of the flux insertion, $\phi$ is adiabatically increased from 0 to $2\pi$ in the DMRG calculations. 
One thereafter obtains the Hall conductance 
$\sigma_H = \frac{e^2}{h}\Delta Q$ by measuring 
the net charge pumping $\Delta Q$ from one edge 
of the cylinder to the other. In DMRG, 
we calculate the net charge transfer as 
$\Delta Q=\sum_{x=L-l+1}^L [\tilde{n}_{x}^{\rm col}(\phi) - \tilde{n}_{x}^{\rm col}(0)]$, 
i.e. the pumped charge to the rightmost $l$ columns
({chosen as $l=3$-4 in practice)} 
where $\tilde{n}_{x}^{\rm col}(\phi)$ 
is the deviation of the charge number of the $x$-th 
column measured in the $\phi$-flux inserted ground state 
$|\psi_\phi\rangle$ from the half filling. For instance, 
we have $\tilde{n}_x^{\rm col}(\phi) = \sum_{y = 1}^W 
\sum_{\lambda = 1}^2 \langle \psi_\phi | \hat{n}_{(x -1)\mathbf{L_1} 
+ y(\mathbf{L_1} - \mathbf{L}_2) + \delta_{\lambda}} - \frac12 | \psi_\phi \rangle$ 
for the XC geometry, and similar expressions for YC.

As shown in \Fig{Fig:Flux}(a), for both XC and YC systems (with widths $W=4$ and 6)
in the QAH phase (e.g., $\alpha=0.15$), we find a net charge 
transfer $|\Delta Q|=1$ through a $2\pi$ flux insertion, 
showing that the Chern number $C=\pm1$. {In addition,} 
\Fig{Fig:Flux}(b) shows the column charge distribution 
$\tilde{n}_{x}^{\rm col}$, where a half-charge $\pm\frac{1}{2}$ 
appears in two edges in $|\psi_{\phi=0} \rangle$.
As $\phi$ gradually increases, the left/right-end charge smoothly 
reduces/increases from $\pm\frac{1}{2}$ to $\mp\frac{1}{2}$, 
which corresponds to an end-to-end pumping of a unit charge 
$\Delta Q=1$, without ``disturbing" the charge distribution in the bulk.
We note that there is two-fold degenerate QAH ground state
(apart from the additional degeneracy due to half-charge 
zero edge modes, see discussion below), 
and the charge pumping could be $\Delta Q= \pm 1$, 
corresponding to the spontaneous TRS breaking states with $C=\pm1$.

\bigskip\noindent\textbf{Understanding the TMI phase.} With DMRG calculations,
we can also calculate the single-particle Green's function 
$G_{\lambda,\lambda'}({\bf R} - {\bf R'}) = \langle c^\dagger_{{\bf R} 
+ \delta_\lambda} c^{\,}_{{\bf R}' + \delta_{\lambda'}} \rangle$, 
from which we can find the electron occupation 
$n_{\lambda,\lambda'}({\bf k})$ in the momentum space. 
Due to the two-sublattice structure, 
$G_{\lambda,\lambda'}({\bf R} - {\bf R'})$ and its Fourier transformation 
$\tilde{G}_{\lambda,\lambda'}({\bf k})$ are both 2$\times$2 matrices 
(cf., Supplementary Note 1). The two eigenvalues 
$\{n_1({\bf k}), n_2({\bf k})\}$ of $\tilde{G}(\bf k)$ are shown in \Fig{Fig:Flux}(d). 
We find for all allowed $\bf k$ points, the larger eigenvalue 
$n_2(\mathbf{k}) \simeq 1$ and the smaller value $n_1(\mathbf{k}) \simeq 0$,
representing the ``two-orbit" electronic structure 
with one orbit filled while the other left empty. 
Albeit small, charge fluctuations between the two orbits are 
still present. We compute the von Neumann entropy 
$S_\mathrm{vN}({\bf k})\equiv-\sum_{\lambda = 1}^2 n_{\lambda}({\bf k})\ln n_{\lambda}({\bf k})$ 
that measures the deviation of the DMRG ground state from a Slater determinant of Bloch states.
In \Fig{Fig:Flux}(e), we show the calculated $S_\mathrm{vN}$ 
averaged over the first BZ, which decreases as $\alpha$ increases, 
and becomes very small for large $\alpha$ cases. For example,
we show the detailed $\bf k$-dependent profile for the $\alpha=0.25$ case, 
in the inset of \Fig{Fig:Flux}(e). The relatively small $S_\mathrm{vN}$ values suggest the QAH state, 
emerging in the interacting TBG model as revealed by DMRG calculations, 
actually very much resembles the Slater determinant ground state of the Haldane 
model and thus can be captured by a mean-field description.

To be specific, for small $\alpha$, a second-order perturbation shows the charging term $\sum_{\varhexagon} ( Q_{\varhexagon} - 1 )^2$ favors the insulating phases in which each hexagon of the honeycomb lattice contains exactly one electron, i.e.~$Q_{\varhexagon} = 1$ for every hexagon. Among all the states satisfying this requirement, the first- and second-order corrections from the cross terms $T_{\varhexagon} (Q_{\varhexagon} - 1)$ vanish. The stripe phase is selected from such states because it minimizes the contribution of $\langle \sum_{\varhexagon} T_{\varhexagon}^2 \rangle$, with the energy $\langle H \rangle_{\rm stripe} \approx \alpha^2 U_0$ 
(c.f. Supplementary Note 2). 

For large $\alpha$, motivated by the resemblance of the DMRG ground state to the Slater determinant, we perform a variational mean-field calculation that approximates the true ground state with the ground state of a tight binding model containing various hoppings (see Methods and Supplementary Note 3). In particular the \Fig{Fig:Model}(g) demonstrates the emergence of NNN currents which
	constitute a loop in each hexagon, spontaneously choosing 
	either the left- or right-chiral direction (here the right chirality). We find that the cross terms, i.e. $\langle T_{\varhexagon} (Q_{\varhexagon} - 1) \rangle_{\text{QAH}}$ become negative and thus favor the QAH phase. Including both the charging terms and $\sum_{\varhexagon} T_{\varhexagon}^2$, the variational mean field analysis results in $\langle H \rangle_{\rm QAH} \approx U_0 (0.037 - 0.27 \alpha + 0.71 \alpha^2)$. Therefore, as $\alpha$ continuously increases from $0$, the mean-field theory also finds the first-order phase transition from the stripe phase to the QAH, in agreement with the DMRG result mentioned earlier. 
The mean-field energy is shown in \Fig{Fig:Phases}(a) as indicated by the blue and red dashed line for the stripe and QAH phases respectively. Both lines provide a good approximation to the DMRG energy curve, and the intersection of two mean-field energies also provides a very good estimate of the QPT value $\alpha^{\rm MF}_c \simeq 0.125$. Interestingly, the energy difference between the mean-field approximation 
and the DMRG calculation decreases as $\alpha$ moves away from the QPT, reflecting the suppression of the quantum fluctuations for large $|\alpha - \alpha_c|$, also illustrated by the $S_\mathrm{vN}$ in \Fig{Fig:Flux}(e). 

Moreover, as shown in Fig.~\ref{Fig:Flux}(b), 
there exist half-charge zero modes 
on both edges of the cylinder with even $W$, 
which also coincide with the Haldane model wrapped on the cylinder 
(for more details, see the Supplementary Note 4). 
We also compute the entanglement spectrum (ES), 
defined as $\xi_i \equiv -\ln(\rho_i)$ 
with $\rho_i$ the eigenvalues of the reduced density matrix. 
As shown in \Fig{Fig:Flux}(c), when we cut at the  
center of the system, 
a two-fold degeneracy in the ES is evident, 
which accounts for the half-charge zero modes 
in the edge [c.f. \Fig{Fig:Flux}(b)], 
through the bulk-edge correspondence.

\bigskip \noindent {\bf Discussion}\\ 
As we mentioned, the QAH can be obtained from narrow band models of TBG with large Coulomb interactions, 
but these models are built in the basis of extended states~\cite{bultinck2019ground,liu2019correlated,
kang2020nonabelian,soejima2020efficient,lian2020tbg} making the interaction potential rather unwieldy. 
The results indeed show that several phases: QAH, strongly correlated topological semimetal, 
and insulating stripe phases, are energetically competitive for the ground states at odd integer fillings
~\cite{kang2020nonabelian,liu2018pseudo,soejima2020efficient,Bernevig2020arXiv,Kwan2021}.

The common belief, however, is that the nontrivial symmetry-protected topology of the narrow bands prevents a faithful construction of models 
within exponentially localized basis even when the bands' total Chern number vanishes~\cite{Po2018}. On the other hand, 
as first shown in the context of the $Z_2$ topological insulators~\cite{Soluyanov2011}, the obstruction is not as severe 
as in the case of a nonzero Chern band (or band composite). If the total Chern number vanishes, 
the exponentially localized Wannier states can be constructed~\cite{Marzari2007}, but some of the protecting symmetries 
do not have a simple on-site implementation~\cite{Marzari2012,Soluyanov2011,XYWang2020}. 
Because the transformation from the Bloch to Wannier basis is unitary and no information is lost in the process, 
it is therefore expected that the lattice tight-binding description should also result in the same ground state 
as found in unobstructed, extended states, basis. However, any practical implementation of this program 
needs to truncate the expansion of the interaction to on-site and few nearest neigbour sites. 
What is not obvious, therefore, is whether all the terms need to be included in the expansion 
or whether it can be truncated to recover the ground state. 

The results presented here show that the truncation at just the nearest neighbor,
parameterized by $\alpha$ in Eq.~(\ref{Eq:H}), is sufficient to recover the insulating and the topologically nontrivial phases. In addition, the main features of the single particle 
excitation dispersion of the strong coupling correlated ground states at the charge 
neutrality point~\cite{YDLiao2020} from the model in Eq.~(\ref{Eq:H}) match those 
computed exactly in the extended basis~\cite{VafekKang2020,Bernevig2020arXiv}. 
This demonstrates the practicality of Wannier description even for such symmetry-obstructed bands. Our real-space interaction-only model therefore establishes the microscopic mechanism 
of the evolution between the insulating stripe and QAH phases. {Our effective model and its unbiased numerical solution therefore revealed the essence of the physics in this particular regime, and is also consistent with other theoretical calculations~\cite{liu2019correlated,kang2020nonabelian,soejima2020efficient,Kwan2021}.}

{As for relevance of our model towards the real system, it is understood that other than the $Q_{\varhexagon}$ and $T_{\varhexagon}$ 
terms, we do not include all the other projected interactions nor the small kinetic terms, i.e., the detailed feature of the TBG material, which will surely modify the specific value of $\alpha_c$. Apart from that they should not qualitatively alter the two phases and 
thus also the main conclusion of the present work. In addition to the ground states 
given above, the dispersion of the charged excitations produced by Eq.~(\ref{Eq:H}) 
is also found to be qualitatively consistent with more detailed calculation by two of the authors 
in Refs.~\cite{VafekKang2020,VafekKang2021Prep}. Ref.~\cite{VafekKang2021Prep} 
has also explicitly shown that the dispersion at the charge neutrality point is dominated by the 
$\alpha$ term in the chiral limit. For systems away from the chiral limit, it is expected that the 
inclusion of other terms may only quantitatively change the dispersion.}


\bigskip \noindent {\bf Methods}\\
\textbf{Density matrix renormalization group.} 
We employ the DMRG method, 
realized in the matrix product state form and with 
U(1) charge symmetry implemented, 
to accurately find the ground state of the TBG model.
Following standard 2D DMRG calculations, 
we map the cylindrical geometries
through a snake-like path, i.e., a quasi-1D structure,
where highly controllable and efficient simulations can be performed. 
In practice, we retain up to $D=512(1024)$ for $W=4(6)$ cylinders, 
with truncation errors $\epsilon<5\times10^{-5}$, 
for an accurate large-scale calculations. 
The detailed convergence check of the TBG model calculations 
can be seen in the Supplementary Note 1.

\noindent \\
\textbf{Mean-field analysis.}
We also applied the mean-field theory to approximate the interactions by a tight binding model with variational hopping constants. The hopping amplitudes are obtained by minimizing the expectation value of the interactions in \Eq{Eq:H} for the state produced by the tight binding model. In practice, the tight binding model includes hopping amplitudes up to the 5th nearest neighbor. The details are presented in the Supplementary Note 3. \\

\section*{Data availability} 
The data that support the findings of this study are available from the corresponding author upon reasonable request. \\

\section*{Code availability} 
All numerical codes in this paper are available upon request to the authors. \\

\section*{Acknowledgements}
B.B.C. and W.L. are indebted to Shou-Shu Gong, Xian-Lei Sheng, 
Xu-Tao Zeng, and Tao Shi for stimulating discussions.
Y.D.L. and Z.Y.M. acknowledge the RGC of Hong Kong SAR of China
(Grant Nos. 17303019, 17301420 and AoE/P-701/20), MOST
through the National Key Research and Development Program
(Grant No. 2016YFA0300502) and the Strategic Priority Research 
Program of the Chinese Academy of Sciences (Grant No. XDB33000000). 
B.-B.C., W.L., and Z.C. acknowledge the support from the NSFC
through Grant Nos. 11974036, 11834014, 12074024, and
11774018. O.V. was supported by NSF DMR-1916958, 
and by the National High Magnetic Field Laboratory through NSF 
Grant No. DMR-1157490 and the State of Florida. 
J.K. acknowledges the support from the NSFC Grant No.~12074276,
and Priority Academic Program Development (PAPD) 
of Jiangsu Higher Education Institutions. We thank the Center for 
Quantum Simulation Sciences at Institute of Physics, 
Chinese Academy of Sciences, the Computational Initiative at the 
Faculty of Science and Information Technology Service at the 
University of Hong Kong, {the HPC Cluster of ITP-CAS,}
and the Tianhe platforms at the National Supercomputer
Centers in Tianjin and Guangzhou for their technical
support and generous allocation of CPU time.

\section*{Author contributions}
Z.Y.M., W.L., and J.K. initiated the work. B.-B.C. and Y.D.L. performed the DMRG 
calculations. J.K. and O.V. conducted the theoretical analysis and mean-field calculations.
All authors contributed to the analysis of the results. W.L., Z.C. and Z.Y.M. supervised the project.

\section*{Additional information}
\noindent
\textbf{Supplementary Information} is available in the online version of the paper. \\
\noindent
\textbf{Competing interests:} The authors declare no competing interests. \\

%

\newpage\clearpage
\onecolumngrid

\renewcommand{\Eq}[1]{Supplementary Eq.~\eqref{#1}}
\renewcommand{\Fig}[1]{Supplementary Figure~\ref{#1}}
\renewcommand{\Tab}[1]{Supplementary Table~\ref{#1}}

\begin{center}
\textbf{\large Supplementary Information for: \\Realization of Topological Mott Insulator in a Twisted Bilayer Graphene Lattice Model}\\
Chen \textit{et al}.
\end{center}

\date{\today}

\setcounter{section}{0}
\setcounter{figure}{0}
\setcounter{equation}{0}
\renewcommand{\thesection}{\Alph{section}}
\renewcommand{\theequation}{\arabic{equation}}
\renewcommand{\thefigure}{\arabic{figure}}

\makeatletter
\renewcommand{\fnum@figure}{Supplementary Figure \thefigure}
\makeatother

\makeatletter
\renewcommand{\fnum@table}{Supplementary Table \thefigure}
\makeatother

\begin{center}
\textbf{Supplementary Note 1: DMRG Results} \\
\end{center}

\begin{figure}[h!]
\includegraphics[angle=0,width=\linewidth]{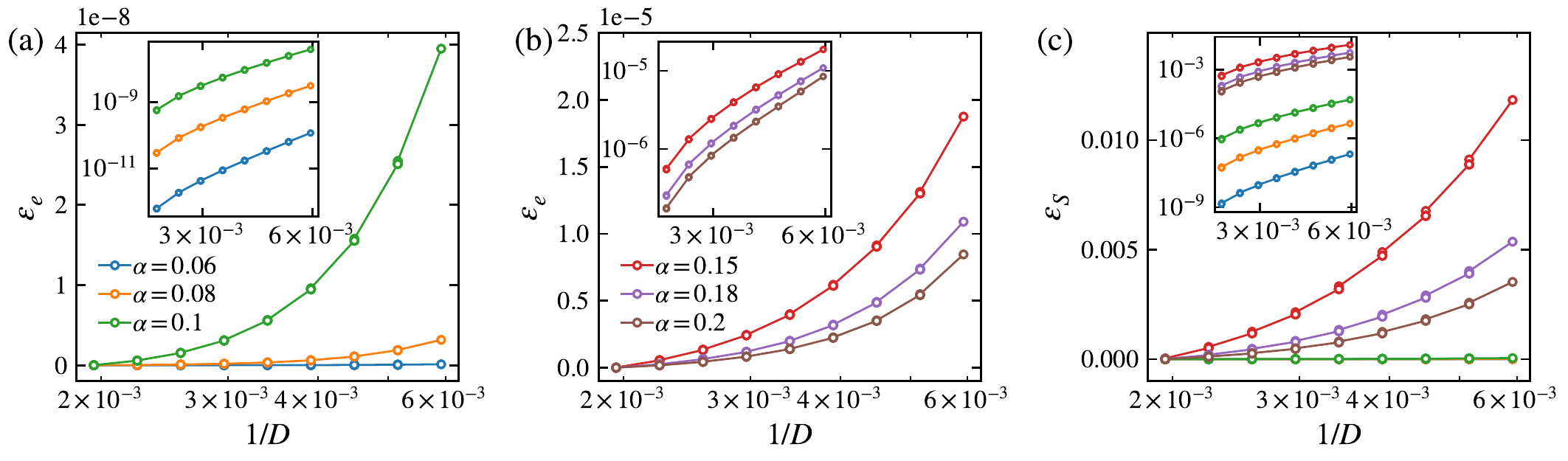}
\caption{\textbf{DMRG convergence check.}
(a) Energy differences between the intermediate sweep $I$ and 
the final sweep $I_f$, $\epsilon_e^{\,}=e(I) - e(I_f)$ as a function of $1/D$ 
in the stripe phase. 
The inset shows $\epsilon_e^{\,}$ in a logarithmic scale.
(b) Same layout as panel (a) but in the QAH phase.
(c) Entanglement entropy differences between the intermediate sweep $I$ and 
the final sweep $I_f$, $\epsilon_S^{\,}=S_E(I) - S_E(I_f)$ as a function of $1/D$ 
in the stripe phase. 
The inset shows $\epsilon_S^{\,}$ in a logarithmic scale.}
\label{Fig:Conv}
\end{figure}

In this section, we present more DMRG results 
further supporting the conclusion in the main text. \\

\textbf{DMRG data convergence.}
Firstly, we show in 
\Fig{Fig:Conv} the computed energy and 
entanglement entropy vs. bond dimension $1/D$. 
in our DMRG calculations. In practice, 
to ensure convergence of the data, 
we ramp up the bond dimensions $D$
in the course of optimization, i.e.,
$D(I) = D(0) \cdot a^I$, where $a$ is a parameter
controlling the increasement of $D(I)$ in the 
$I$-th step, with $I \in 0,1,\cdots, I_f$. 
To be specific, starting with an initial bond dimension $D(0)$, 
we increase the bond dimension $D(I)$, step by step,
until the final bond dimension $D(I_f)$ is reached. 
In practical calculations, we set $D(0)=128$, $a=2^{1/5}$, 
and $D(I_f)=512$($1024$) for width $4$($6$) cylinder, the results are very well converged. 
In addition, at each intermediate step $I$, 
we perform $5$ sweeps before moving to the 
next step $I+1$ with increased bond dimension $D(I+1)$.

In \Fig{Fig:Conv}(a,b), we show the differences 
of the calculated ground-state energy 
$\epsilon_e^{\,}=e(I) - e(I_f)$ versus the retained bond dimension 
$1/D$, from which one can observe that, 
for both stripe [panel(a)] and QAH [panel (b)] phases, 
the energy have well-converged within $\epsilon_e^{\,}\sim10^{-12 \sim -9}$ 
and $10^{-7\sim-6}$, respectively. The entanglement entropy 
differences $\Delta S_E$ are shown in \Fig{Fig:Conv}(c), 
which also show good convergence, 
with $\epsilon_S^{\,} \sim 10^{-9\sim-3}$, 
depending on the specific $\alpha$ parameters.\\

\begin{figure}[h!]
\includegraphics[angle=0,width=0.7\linewidth]{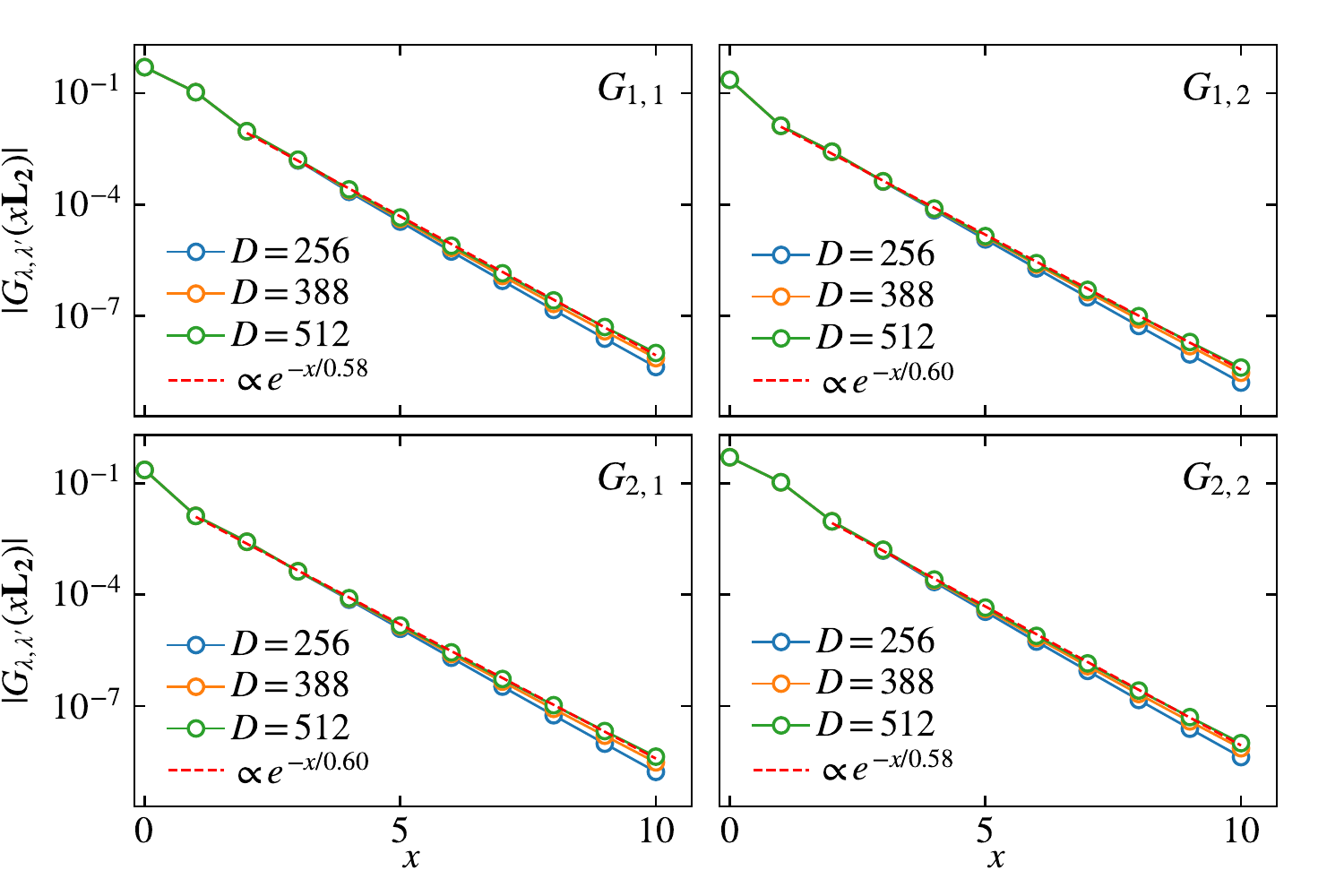}
\caption{\textbf{Single-particle Green's function.}
In a YC$4\times24\times2$ system with $\alpha=0.15$ 
(QAH phase), single-particle Green's functions 
$G_{\lambda,\lambda'}(x{\bf L_2})$ are calculated, 
which are shown to be well converged vs. $D$.
The four components $G_{1,1}, G_{1,2}, G_{2,1}, G_{2,2}$ 
are plotted versus $x$ in (a-d) panels, which all
decay exponentially as $\propto e^{-x/\xi}$, 
with similar correlation lengths $\xi\simeq0.6$.
}
\label{Fig:GF}
\end{figure}

\textbf{Single-particle Green's function.} 
In \Fig{Fig:GF} we show the single-particle
Green's function results on YC4 geometry, 
where $G_{\lambda,\lambda'}(x{\bf L_2}) = \langle c^\dagger_{{\bf R} + \delta_\lambda} c^{\,}_{{\bf R} + x{\bf L_2} + \delta_{\lambda'}} \rangle$ 
is computed by DMRG, with 
${\bf R}= \frac{W}{2}{\bf (2L_1-L_2)}$.
From \Fig{Fig:GF}, we find a very clear 
exponential decay of all four elements 
of the $2\times 2$ Green's function matrix,
with a rather short correlation length $\xi<1$ 
(in the unit of NN edge of honeycomb lattice).
These results are consistent with the flat 
entanglement entropy shown in the inset 
of Fig.~2 (b) in main text, 
pointing to a ground state with nonzero charge 
gap in the bulk.\\ 

\begin{figure}[h!]
\includegraphics[angle=0,width=0.8\linewidth]{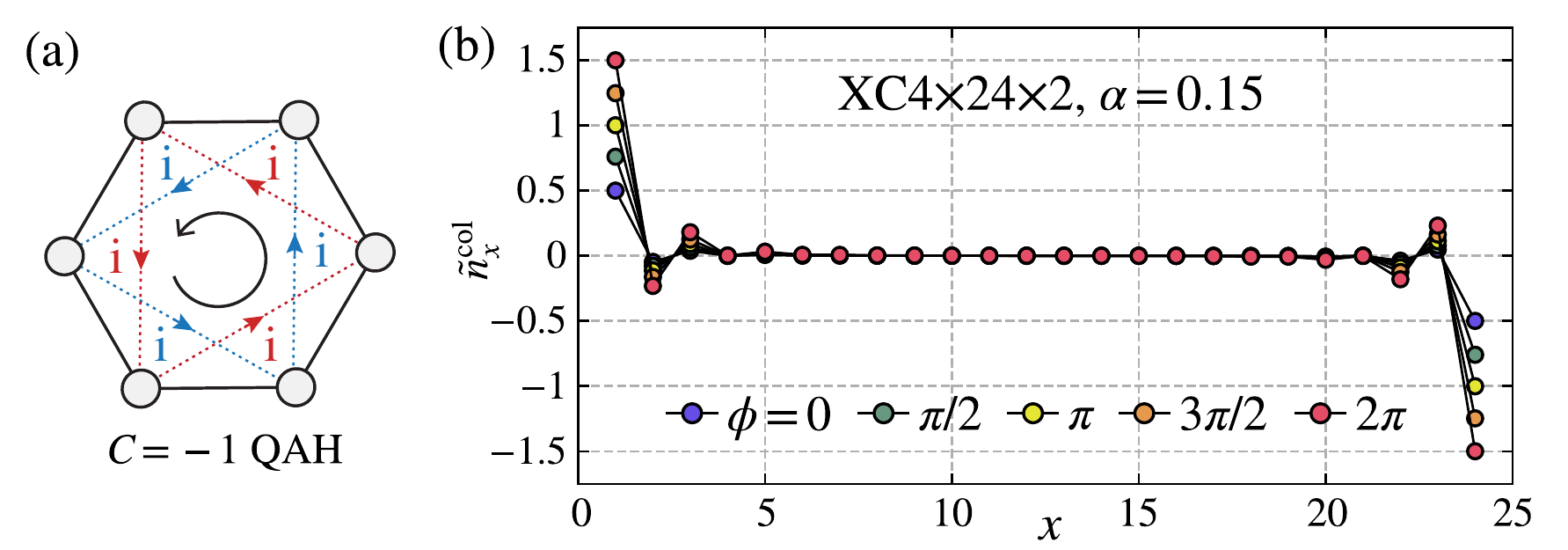}
\caption{\textbf{Real-space charge distributions.}
(a) The rotation of loop current is opposite with the one ($C=1$ QAH) in main text 
[c.f. inset of Fig.~1(e)], thus a $C=-1$ QAH state.
(b) In a XC$4\times24\times2$ system with $\alpha=0.15$ 
(QAH phase), the charge distributions for different 
$\phi$-flux threading the cylinder hole. An electron is pumped
from the right edge to the left, resulting $\pm\frac{3}{2}$ 
charges in both end, corresponding to a $C=-1$ QAH state.
}
\label{Fig:QAHm}
\end{figure}

\textbf{QAH state with Chern number $C=-1$.} 
Since the QAH state in the large-$\alpha$ phase
spontaneously breaks the time-reversal symmetry,
the topological states can thus have Chern numbers 
$C=1$ and $-1$. In \Fig{Fig:QAHm} we show 
the ground state with $C=-1$, which have roughly 50\% 
probability to appear in our calculations. 
From \Fig{Fig:QAHm}(a), 
we find the directional circular currents, 
signaling the TRS breaking, follow rightly the
opposite chirality to the QAH state shown in 
Fig.~1(e) of the main text where the $C=1$ QAH state is realized. Another
distinction is the fractional charge zero modes
on the edges of cylinders, upon flux insertion.
For the $C=-1$ state, when we thread a flux
from left to right [following exactly the inset of 
Fig.~3(a) in main text], we find an unit charge 
$\Delta Q=1$ is pumped from right edge to the 
left. This is revealed explicitly in \Fig{Fig:QAHm}(b), where we find 
the end charge increases from $1/2$ to $3/2$ on the left edge,
while decreases from $-1/2$ to $-3/2$ on the right edge. \\

\newpage
\begin{center}
\textbf{Supplementary Note 2: Stripe phase with small $\alpha$} \\
\end{center}

In this section, we study the possible ground state of the Hamiltonian Eq.~(1) in main text
with small $\alpha$. Here, we follow the analysis in Ref.~\cite{kang2019strong}. 
The Hamiltonian can be expanded as
\begin{align}
H & = H_0 + 2 \alpha H_1 + \alpha^2 H_2 \\
H_0 & = \sum_{\varhexagon} (Q_{\varhexagon} - 1)^2 \\
H_1 & = \sum_{\varhexagon} T_{\varhexagon}  (Q_{\varhexagon} - 1)  \\
H_2 & =  \sum_{\varhexagon}   \left( T_{\varhexagon} \right)^2
\end{align}
Note that $T_{\varhexagon}$ commutes with $Q_{\varhexagon}$ 
because the hopping of $T_{\varhexagon}$ occurs inside the hexagon, 
and thus does not change the total charges on the six vertices of the hexagon. 
Therefore, $T_{\varhexagon}  (Q_{\varhexagon} - 1) = (Q_{\varhexagon} - 1) T_{\varhexagon}$.
  
When $\alpha$ is small, we can treat $H_1$ and $H_2$ as perturbations with respect to the leading term  $H_0$. At the half filling, the ground state of $H_0$ is given by the constraints $\left( Q_{\varhexagon} - 1 \right)| \Psi_0 \rangle = 0$ for all the hexagons $\varhexagon$ and thus the corresponding energy $E_0 = 0$. This constraint is satisfied by a large manifold of states, including both the stripe and sublattice polarized states. Now, consider the perturbation of $H_1$ and $H_2$ within the manifold of the degenerate states. For any state $\Psi_0$ in this manifold, $\left( Q_{\varhexagon} - 1 \right)| \Psi_0 \rangle = 0$. As a consequence, both the first and second order perturbations of $H_1$ vanishes:
\begin{align}
&  \sum_{\varhexagon} \langle \Psi_0' |  T_{\varhexagon}  (Q_{\varhexagon} - 1)  | \Psi_0 \rangle = 0 \\
&  \sum_{\varhexagon} \sum_{\varhexagon'} \sum_{n \not\in \Psi_0} \frac1{E_0 - E_n} \times \langle \Psi_0' | (Q_{\varhexagon} - 1) T_{\varhexagon}  | n \rangle \langle n | T_{\varhexagon'}  (Q_{\varhexagon'} - 1)  | \Psi_0 \rangle  = 0
\end{align}
where $\Psi_0$ and $\Psi_0'$ are two arbitrary orthogonal states inside the ground state manifold of $H_0$, and $| n \rangle$ labels the excited state of $H_0$. Up to $O(\alpha^2)$, we also need to include the first order correction of $H_2$, i.e., $\sum_{\varhexagon} \langle \Psi_0' | \left( T_{\varhexagon} \right)^2 | \Psi_0   \rangle$.  After expanding the square form into the four-fermion terms, it is obvious that only the following terms
\[  \sum_{\bf R} \sum_{l =1}^6 \sum_{\eta = \pm 1} \sum_{\lambda} c^{\dagger}_{\mathbf{R} + \delta_l, \lambda} c_{\mathbf{R} + \delta_{l + \eta}, \lambda} c^{\dagger}_{\mathbf{R} + \delta_{l + \eta}, \lambda} c_{\mathbf{R} + \delta_l, \lambda}   \] survive, where the index $\bf R$ refers to the hexagon and $\mathbf{R} + \delta_l$ and $\mathbf{R} + \delta_{l \pm 1}$ are two neighboring vertices of this hexagon. These terms do not change the total charge of each hexagon. It is obvious that their first order correction is $1$ if the site $\mathbf{R} + \delta_l$ is occupied and the site $\mathbf{R} + \delta_{l \pm 1}$ is empty, and becomes $0$ otherwise. Therefore, among the states in which $Q_{\varhexagon} = 1$ for every hexagon, this correction is minimized by decreasing number of ``dangling'' bonds that connect an occupied site and an empty one. At the half filling,
the number of such bonds becomes smallest for the stripe phase, 
as shown in Fig.~1(f) of the main text. 
The corresponding first order 
energy correction per site is thus 
\begin{equation}  
 \delta E/N = \alpha^2 U_0,
\end{equation}
and plotted as the blue dashed curve in Fig.~2(a) of the main text. \\

\begin{figure}[tb]
\includegraphics[angle=0,width=0.4\linewidth]{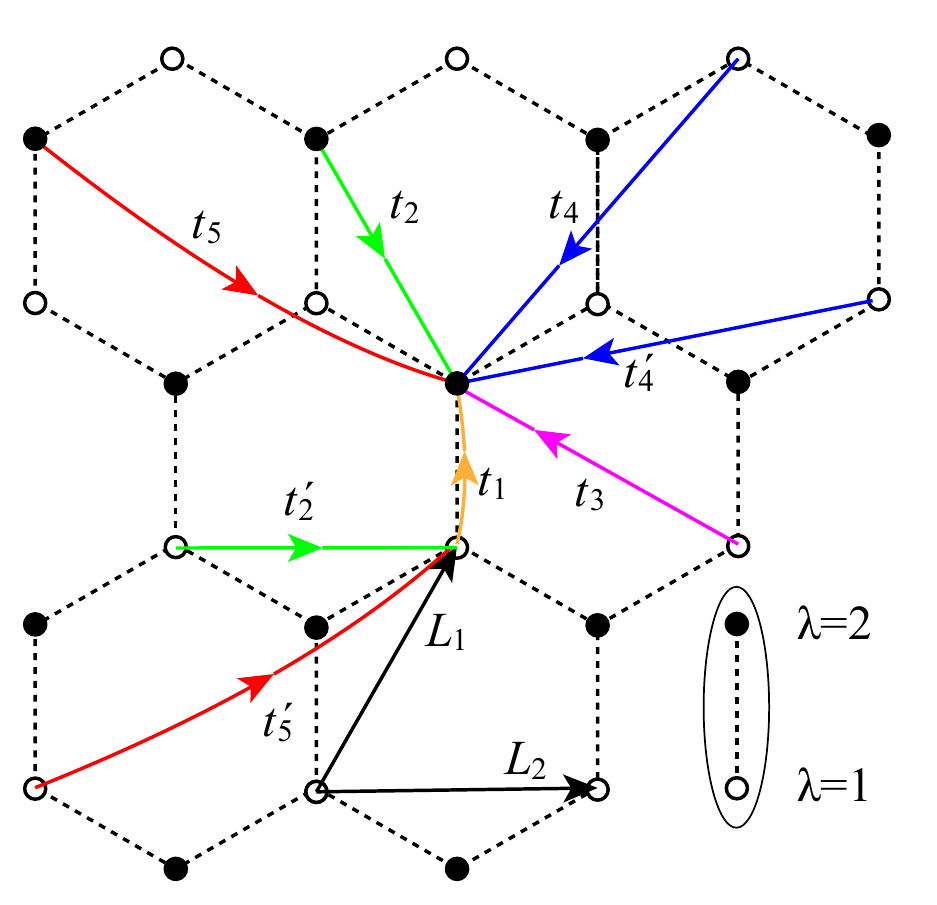}
\caption{ \textbf{Hopping terms in the mean field theory.}
Not all the hoppings are presented, as they can be 
obtained by the $C_3$ rotation symmetry applied on 
the sketched ones.}
\label{FigS:Hopping}
\end{figure}

\begin{center}
\textbf{Supplementary Note 3: Mean Field Approximation with Large $\alpha$} \\
\end{center}
When $\alpha$ becomes larger, the perturbation theory in the previous section fails.
Our DMRG calculation has revealed that the QAH state appears with $\alpha \gtrsim 0.12$ and furthermore, the state can be approximated as the Slater determinant of the Bloch states. Motivated by these DMRG results, we consider a $C_3$ symmetric tight binding model with the hopping constants up to the \textit{fifth} nearest neighbor. 
As shown in \Fig{FigS:Hopping}, the hopping terms are

\begin{eqnarray}
\label{EqS:Hopping}
H_t & & = H_1 + H_2 + H_3 + H_4 + H_5 \\
H_1 & & = \sum_{\bf R} t_1 c^{\dagger}_{\mathbf{R} + \delta_2} \left( c_{\mathbf{R} + \delta_1} +  c_{\mathbf{R} + \mathbf{L}_1 + \delta_1} + c_{\mathbf{R} + \mathbf{L}_1 - \mathbf{L}_2 + \delta_1} \right) + h.c.  \nonumber  \\
H_2 & & = \sum_{\bf R} t_2  c^{\dagger}_{\mathbf{R} + \delta_2}   \left(  c_{\mathbf{R} + \mathbf{L}_1 - \mathbf{L}_2 + \delta_2} +  c_{\mathbf{R} +  \mathbf{L}_2 + \delta_2} + c_{\mathbf{R}  - \mathbf{L}_1 + \delta_2} \right) + h.c. \nonumber \\
& & \quad +   t_2' c^{\dagger}_{\mathbf{R} +  \delta_1}   \left(  c_{\mathbf{R}  + \mathbf{L}_2 - \mathbf{L}_1, 1} +  c_{\mathbf{R} - \mathbf{L}_2 + \delta_1} + c_{\mathbf{R} + \mathbf{L}_1 + \delta_1} \right)  + h.c. \nonumber \\
H_3 & & =\sum_{\bf R}  t_3  c^{\dagger}_{\mathbf{R} + \delta_2} \left(  c_{\mathbf{R} + 2\mathbf{L}_1 - \mathbf{L}_2 + \delta_1} +  c_{\mathbf{R} + \mathbf{L}_2 +  \delta_1} + c_{\mathbf{R} - \mathbf{L}_2 + \delta_1} \right) + h.c.  \nonumber  \\
H_4 & & = \sum_{\bf R}  t_4 c^{\dagger}_{\bf{R} + \delta_2}   \left(  c_{\mathbf{R} + \mathbf{L}_1 + \mathbf{L}_2 + \delta_1} +  c_{\mathbf{R} + 2\mathbf{L}_1 - 2 \mathbf{L}_2 + \delta_1} + c_{\mathbf{R} - \mathbf{L}_1 + \delta_1} \right) + h.c. \nonumber \\
& & \quad + t_4' c^{\dagger}_{\mathbf{R} + \delta_2}   \left(  c_{\bf i + \bf 2L_1, 1} +  c_{\bf i + \bf L_1 - 2\bf L_2, 1} + c_{\bf i + \bf L_2 - \bf L_1, 1} \right) + h.c. \nonumber \\
H_5 & & = \sum_i t_5  c^{\dagger}_{\bf i, 2}   \left(  c_{\bf i + \bf L_1 + \bf L_2, 2} +  c_{\bf i  + \bf L_1 - 2\bf L_2, 2} + c_{\bf i + \bf L_2 - 2\bf L_1, 2} \right) + h.c. \nonumber \\
& & \quad +  t_5' c^{\dagger}_{\bf i, 1}   \left(  c_{\bf i + 2\bf L_1 - \bf L_2, 1} +  c_{\bf i - \bf L_2 - \bf L_1, 1} + c_{\bf i + 2\bf L_2 - \bf  L_1, 1} \right)  + h.c. \nonumber 
\end{eqnarray}
where $H_{l}$ ($l = 1, \cdots, 5$) are the hopping terms between nearest neighbors, next nearest neighbors, ...., up to the 5th nearest neighbors, respectively. The index $\bf R$ labels the position of unit cell, 
and $\lambda =1$ (or $2$) is the index of the sublattices.

This tight binding model can be diagonalized in the momentum space, and the chemical potential $\mu$ is fixed by the particle number. The tight binding model produces the many-body ground state as
\[ | \psi \rangle = \prod_{\substack{i, \bf k \\ E_i(\bf k) < \mu}} d^{\dagger}_i(\bf k) | \emptyset \rangle  \]
where $i = 1$ or $2$ labels the two bands of the tight binding model $H_t$. Applying the variational mean field approximation, we minimize $E_{\mathrm{mf}} = \langle \psi | H | \psi \rangle$ with respect to all the hopping parameters in $H_t$. Here, $H$ is the interaction Hamiltonian in Eq.~(1) of the main text.

For all the possible values of $\alpha$, we numerically found out that all the hopping parameters are purely imaginary and $t_2 = t_2'$. Furthermore, the hoppings beyond the 3rd nearest neighbor are tiny, and thus can be neglected in $H_t$.  We have found that $E_{\mathrm{mf}}$ is minimized only when the hoppings of $H_t$ are imaginary and thus lead to the QAH state. Therefore, $H_t$ is qualitatively similar to the Haldane model up to a $U(1)$ gauge transformation.

The expectation value $E_{\mathrm{mf}}$ is plotted as the red dashed
curve in Fig.~1(a) of the main text. The relative
difference between $E_{\mathrm{mf}}$ and the energy produced by
DMRG becomes smaller as $\alpha$ increases. Our variational mean
field calculations suggest that this difference originates from the quantum
fluctuation, that also becomes smaller as $\alpha$ increases as 
suggested in  Fig.~3(e) of the main text. 

\begin{table}[tb]
\caption{\textbf{Variational parameters and the Green's function}. For $\alpha=0.15$ case, 
the variational parameters of hopping amplitudes and the Green's function 
obtained from both mean-field and DMRG calculations are listed up to 3rd nearest neighbor.}
\begin{tabular}{cccc}
    \toprule
     &\hspace{8pt} varitional parameters & \hspace{8pt}Green's function (mean-field)  &\hspace{8pt} Green's function (DMRG) \\
    \hline
    NN & $t_1/U_0\simeq0.2i$& $0.232i$ & $0.224i$  \\
    NNN &$t_2/U_0\simeq0.116i $&$ 0.106i$& $0.099i$ \\
    $3^{rd}$ NN &$t_3/U_0 \simeq-0.102i$& $-0.080i$ & $-0.074i$ \\
     \hline\hline
\end{tabular}
\label{TabS:Hopping}
\end{table}

Besides the variational energy, we have also compared the single-particle
Green's functions obtained from mean-field calculations and DMRG. 
As shown in \Tab{TabS:Hopping}, the hopping amplitudes (as variational parameters 
in the mean-field calculations) are listed, up to 3rd nearest neighbor in the first column. 
The corresponding single-particle Green's functions from both the mean-field and 
DMRG calculations show excellent agreement, with differences $<0.01i$, 
confirming the effectiveness and accuracy of the mean-field theory in understanding the 
QAH phase in the interaction-only TBG superlattice model.\\

\begin{center}
\textbf{Supplementary Note 4: Half-charge Zero Modes on the Edge} \\
\end{center}

\begin{figure}[tb]
\includegraphics[angle=0,width=.8\linewidth]{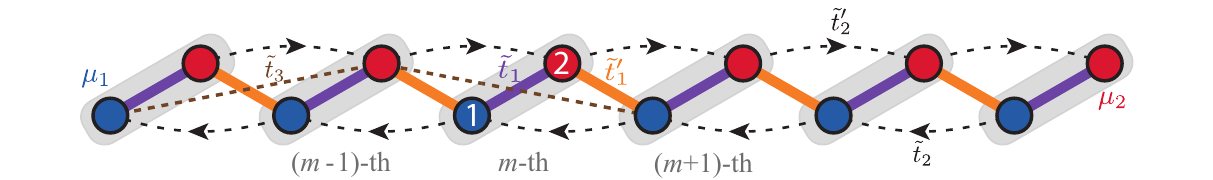}
\caption{\textbf{The generalized SSH model}. Illustration of the 1D model $H_\mathrm{1D}(k)$,
with both NN ($\tilde t_1, \tilde t'_1$) and NNN ($\tilde t_2, \tilde t'_2$) 
hopping amplitudes, as well as sublattice chemical potential $\mu_1$ and $\mu_2$.}
\label{FigS:SSH}
\end{figure}

In this section, we provide a detailed analysis of the particle occupation number and
half-charge zero modes on the edges, based on our tight binding model on a cylinder. 
Here, we focus on the XC geometry with $W$ unit cells along the  periodic direction, so that
\[ c_{ \mathbf{R} + \delta_\lambda} \equiv c_{ \mathbf{R} + W \mathbf{L}_2 + \delta_\lambda}  \] 
where $\lambda = 1$ or $2$ refers to the sublattice.
For the sake of simplicity, in the following discussions, we only keep 
$H_t$ [c.f. \Eq{EqS:Hopping}] up to the 
\textit{third} order, denoted as $H_t^{(3)}$. 
Rewriting $H_t^{(3)}$ through Fourier transformation along $\mathbf{L}_2$ direction 
\[ c_{m, \lambda}(k) = \frac1{\sqrt{W}} \sum_{l = 1}^W c_{l \mathbf{L}_1 + m \mathbf{L}_2 + \delta_\lambda} e^{- i k l} \ , \]
we arrive at the Hamiltonian in the hybridized $(m, k)$ space, 
\begin{eqnarray} \label{EqS:extSSH}
H_t^{(3)} & = & \sum_{m,k}\Big[t_1(1 + e^{i k})c^\dagger_{m,2}(k) c_{m,1}(k)
+ t_1 e^{i k} c^\dagger_{m,2}(k) c^{\,}_{m - 1,1}(k) \\
\phantom{\sum_i} & +\,& t_2 e^{-i k}c^\dagger_{m,2}(k) c^{\,}_{m,2}(k)
+ (t_2^* e^{-i k} + t_2)\,c^\dagger_{m,2}(k) c^{\,}_{m + 1, 2}(k) \nonumber\\
\phantom{\sum_i} & +\,& t_2 e^{i k} c^\dagger_{m,1}(k) c^{\,}_{m,1}(k) 
+ (t_2^\ast  + t_2 e^{-i k})\, c^\dagger_{m,1}(k) c^{\,}_{m+1,1}(k) \nonumber\\
\phantom{\sum_i} & +\,& t_3(1 + e^{2 i k})\, c^\dagger_{m,2}(k) c^{\,}_{m - 1,1}(k) + t_3 c^\dagger_{m,2}(k) c^{\,}_{m + 1, 1}(k) \Big]+ h.c.\nonumber
\end{eqnarray}
We recognize $H_t^{(3)}$ as a summation of $W$ 
decoupled 1D chains, i.e., $H_t^{(3)} = \sum_{k} H_\mathrm{1D} (k)$ with
\begin{eqnarray}
H_\mathrm{1D} (k) & = & \sum_m (\tilde t_1 c^\dagger_{m,2}(k) c^{\,}_{m,1}(k)
+ \tilde t_1' c^\dagger_{m,2}(k) c^{\,}_{m + 1, 1}(k) + \tilde t_3 c^\dagger_{m,2}(k) c^{\,}_{m - 1, 1}(k) ) +h.c. \nonumber \\
& + & \sum_{m}  (\tilde t_2 \,c^\dagger_{m,1}(k) c^{\,}_{m+1,1}(k) + 
\tilde t_2' \,c^\dagger_{m,2}(k) c^{\,}_{m+1,2}(k))  + h.c. \nonumber\\
& + & \sum_{m,\lambda} \mu_\lambda c^\dagger_{m,\lambda} c^{\,}_{m,\lambda} \ .
\end{eqnarray}
This 1D generalized SSH Hamiltonian is illustrated in \Fig{FigS:SSH}, where the hopping amplitudes are
\begin{align} 
\tilde t_1 & = t_1(1+e^{i k}) & \tilde t_1'& =t_3 & \tilde{t}_3 & = t_1 e^{i k} + t_3 (1 + e^{2i k}) & \tilde t_2 & = (t_2^* + t_2 e^{-i k}) \\
\tilde t_2' & = (t_2 + t_2^* e^{-i k})  & \tilde \mu_1 & = 2\mathrm{Re}(t_2 e^{i k}) & \mu_2 & =  2\mathrm{Re}(t_2 e^{- i k}).
\end{align}

\begin{figure}[tb]
\includegraphics[angle=0,width=0.9\linewidth]{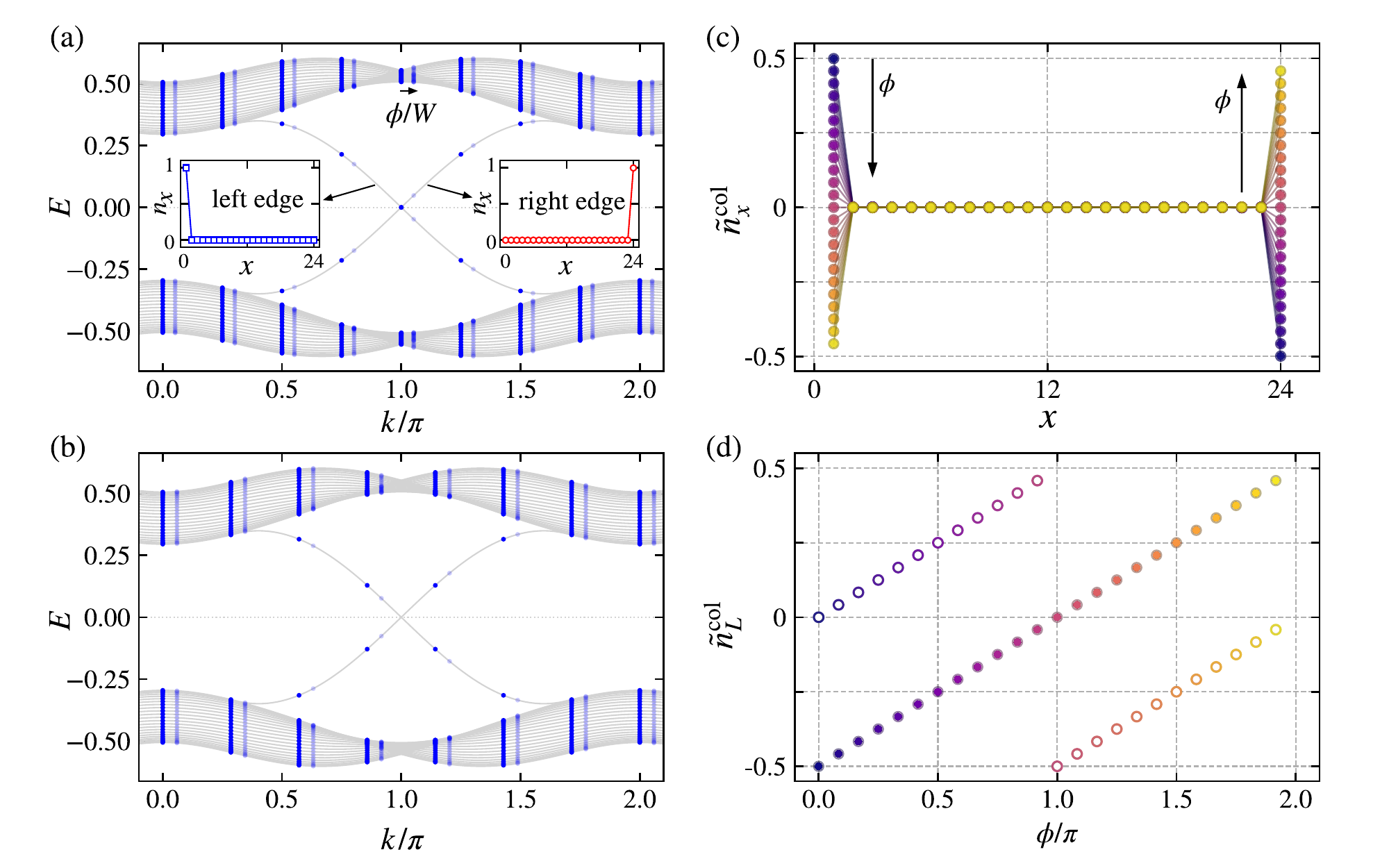}
\caption{\textbf{Flux insertion in the tight-binding model.} 
(a) The spectrum of $H_t^{(3)}$ with even width $W=8$ and 
hopping parameters $t_1=0.2\mathrm{i}$, $t_2=0.116\mathrm{i}$ and 
$t_3=-0.102\mathrm{i}$, taken from the $\alpha=0.15$ DMRG 
calculations. All the allowed momenta are colored dark blue, 
which are shifted to the light-blue dots by $\Delta k=\phi/W$ in the flux insertion case.
The insets show the single-particle charge distributions of the two edge modes.
(b) Same layout as panel (a) with odd width $W=7$ otherwise.
(c) For even width system ($W=8$ here), the charge distribution over column index $i$ 
is shown for flux $\phi\in[0,2\pi)$, with the color code indicated by panel (d). 
(d) Number of charge at the right edge $\tilde n_{i=L}^\mathrm{col}$ for $W=8$(filled) 
and $W=7$(hollow) system.}
\label{FigS:Flux}
\end{figure}

\textbf{Zero modes on the chiral edge at $\pi$ momentum.}
In \Fig{FigS:Flux}(a,b), we show the dispersion of $H_t^{(2)}$ [\Eq{EqS:extSSH}] 
on an open-ended cylinder as a function of $k$, the momentum along the periodic direction. 
While the bulk states are clearly gapped, there exists two branches of chiral edge modes that 
are gapless [c.f., insets of \Fig{FigS:Flux}(a) for their charge distributions localized 
on the edge], corresponding to the bulk QAH topological state. 
Interestingly, the two edge modes are found to be degenerate only at $k = \pi$. 
where the two branches cross. This can be understood as follows, by introducing a sublattice dependent gauge transformation $\tau_z$, 
under which $c_{\mathbf{R} + \delta_l} \longrightarrow (-)^l c_{\mathbf{R} + \delta_l}$.
It is obvious that the interaction $H$ [Eq.~(1) of the main text] is invariant under the 
combined symmetry $\tau_z C_{2z}$, where $C_{2z}$ is the two-fold rotation around the center of a plaquette. Furthermore, the mean field Hamiltonian $H_t^{(3)}$ with all imaginary hoppings is also invariant under $\tau_z C_{2z}$.
As a consequence, the two edge modes must cross at $C_{2z}$ invariant momentum, i.e., at $k = \pi$. This crossing also sets the chemical potential at the half filling if it is inside the insulating gap. For even $W$, the system has to occupy one of the two states at the crossing, so that $|n^{col}_1 - n^{col}_L| = 1$, leading to the appearance of half charges on both sides.
Notably, the conclusion of half-charge zero 
modes at $k = \pi$ holds only for even $W$ since the edge state with $k = \pi$ does not exist if $W$ is odd. \\

\textbf{Flux insertion in the tight-binding model.} 
Below, we stick to the more simpler $H_t^{(3)}$ and analyze the charge pumping therein 
through flux insertion. 
When $W$, the number of unit cell along the periodic direction, is finite, 
the set of all possible momenta is finite, with $k = \frac{2\pi n}{W}$ and 
$n = 0, 1, \cdots, W -1$ if the magnetic flux is absent [see the dark blue dots 
in \Fig{FigS:Flux}(a,b)]. If $W$ is odd, $k$ can never be $\pi$, 
and thus the system only fills all the states below $E = 0$ in \Fig{FigS:Flux}(b). 
Therefore, the two edges have the same number of particles. 
The situation is quite different for even $W$ [c.f. \Fig{FigS:Flux}(a)], 
where the system can only fill one of the two zero modes on each edges. 
Therefore, there is one more particle on one edge than the other, 
leading to the appearance of half-charge zero modes on both edges. 

Furthermore, when flux $\phi$ is inserted, the momentum $k$ shifts by $\phi/W$ 
[see the light-blue dots in \Fig{FigS:Flux}(a,b)]. 
As shown in \Fig{FigS:Flux}(c), for the even $W$ case, as the flux increases 
in a $2\pi$ period, the half charge gradually fades away and disappears for $\phi=\pi$, 
which then reappear with its sign reversed. The net charge transfer $\Delta Q$ from 
the left to the right edge in the course of charge pumping is shown in \Fig{FigS:Flux}(d). 

On the other hand, for cylinders with odd $W$, the system starts from the initial state
with no half-charge zero modes (as $k=\pi$ can not be selected due to the cylinder geometry),
and the pumped charge undergoes a jump right at $\phi=\pi$, when the half-charge zero modes
restore. As $\phi$ exceeds $\pi$, the half-charge zero mode switches to the left end (instead of
the right one), and thus the $\tilde n_L^{\rm col}$ jumps from $+0.5$ to $-0.5$, and then gradually increases
as the flux $\phi$ increases and vanishes again for $\phi=2\pi$. We note that the ``jump" of charge distribution
in the flux insertion procedure happens in the tight-binding model $H_t^{(3)}$ [\Fig{FigS:Flux}(d)] does
not necessarily take place in the adiabatic DMRG simulations of XC$W$ with $W$ odd. 
As there we feed the ground state of previous flux $\phi$ as the initial state of next flux $\phi$, 
and the charge pumping from one end to the other can be realized, without such abrupt ``jump", in practice. 
\\

\begin{figure}[tb]
\includegraphics[angle=0,width=1\linewidth]{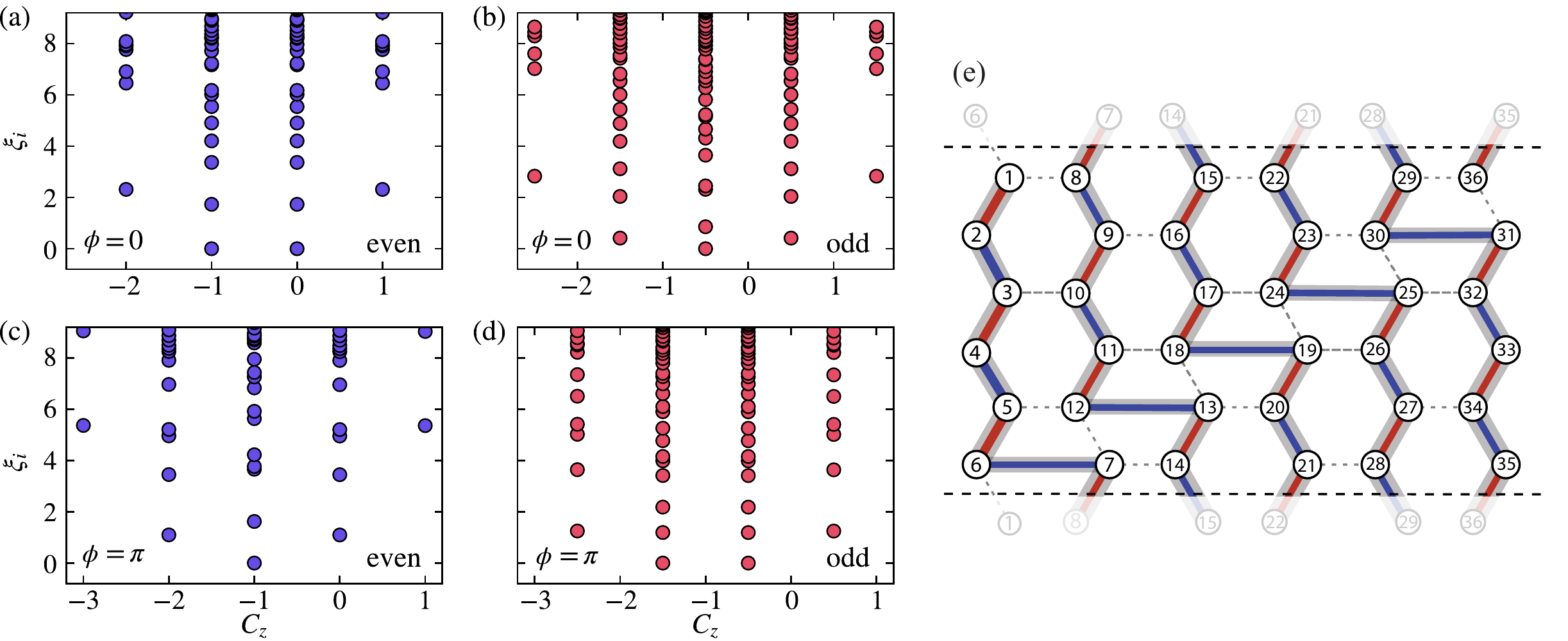}
\caption{\textbf{Degeneracies in the entanglement spectra.} In XC$4\times24\times2$ 
system with $\alpha=0.15$ and $\phi=0$, the entanglement spectra are shown when cut at an 
(a) even and (b) odd bond in the central regime
of the DMRG snake path. (c,d) Same layout as (a,b) otherwise for the $\phi=\pi$ case. 
(e) The snake path of an XC$3\times6\times2$ cylinder is shown 
with site ordering explicitly labeled, and the even/odd bonds are colored with blue/red.
The two horizontal dashed lines indicates the periodic
boundary conditions along the $y$ (vertical) direction of the cylinder geometry.
}
\label{FigS:ES}
\end{figure}

\textbf{Symmetry protected half-charge zero modes and degeneracies of entanglement spectrum.}
This half-charge zero modes on both cylinder edges, which disappear
at the inserted flux $\phi=\pi$, is also reflected in the bulk property,
in terms of the entanglement spectrum degeneracy. As shown in \Fig{FigS:ES}(a,b),
in a XC$4\times24\times2$ cylinder with $\alpha=0.15$ (QAH phase), 
the entanglement spectrum on an even bond [c.f. panel (a)] 
exhibits a two-fold degeneracy, while that on the odd bond is
non-degenerate. As one threads a flux through the cylinder, 
the two-fold degeneracy is lifted. However, and very interestingly, 
at $\phi=\pi$ the two-fold degeneracy reappears on the odd bond (with even one non-degenerate). 
This is in accordance with the absence of zero edge modes at $\phi=\pi$,
as shown in \Fig{FigS:Flux}(b).

\end{document}